\documentstyle[12pt]{article}

\def\tend{\mathop{\to}}

\def\lim{\mathop{\rm {lim}}}

\begin{document}
\large
\begin{center}
\bigskip
{\bf Retardation effects from quark confinement on low-energy nucleon
dynamics }\\
\bigskip
{\rm
Renat Kh.Gainutdinov and Aigul A.Mutygullina}\\\bigskip
\it
Department of Physics\\
Kazan State University,\\
18 Kremlevskaya St, Kazan 420008,\\
Russia\\
E-mail: Renat.Gainutdinov@ksu.ru
\end{center}

\parindent=2.5em

\section*{I. Introduction}

Study of the effects of quarks and gluons confined within hadrons on
nucleon dynamics is of fundamental importance in understanding 
the strong interaction. These effects are important for describing the
short-range (SR) part of the $NN$ interactions. They can also be very
important for describing the dynamics of dense nucleon matter especially
at large densities. As has been shown in Ref.[1], the internal quark structure
of baryons can have substantial effects on the composition and structure of
neutron star matter. Since a nonperturbative treatment of quantum
chromodynamics (QCD) is not possible today, the effects of the internal
quark structure of hadrons on low-energy nucleon dynamics are investigated by
using quark models (see, for instance, Refs.[2-9]). However, 
these models completely miss the effects
of retardation in the gluon-exchange interaction, despite the fact that
they can be important [10], and on the whole the effects of 
quark-gluon retardation on nucleon dynamics are poorly understood at
present. 

The effects of quark-gluon retardation on nucleon dynamics differ
profoundly from the well-known meson-retardation effects that are taken
into account, for example, in the Bonn model [11]. Meson retardation
effects give rise to nonlocality in time of the $NN$ interaction, and hence 
to an energy dependence of the effective potentials describing these
interactions, which can have significant effects on three- and many
nucleon results [12].
Nonlocality in time of such interactions is an expression of a loss
of probability from the two-nucleon subspace of the Hilbert space 
of hadron states. Obviously, 
quark-gluon retardation, that has to be taken into account in describing
hadron dynamics, should also result in nonlocality in time of hadron
interactions. However, due to confinement this must not lead to 
a loss of probability from the Hilbert space of hadron states.
On the other hand, as is well known,
within the Hamiltonian formalism the interaction generating the
dynamics of a quantum system may be nonlocal in time only in the case when
the system is not closed, and as a result the evolution is not unitary.
Similar problems arising due to nonlocality in time of the $NN$
interaction has been discussed in [13-20]. To solve these problems quantum
dynamics need to be extended to describe the
unitary evolution of a closed system in the case where the interaction
generating the dynamics of the system is nonlocal in time. For the first
time, this problem has been solved in Ref.[21], where it has been shown
that the use of the Feynman approach [22,23] to quantum theory in combination
with the canonical approach allows such an extension of quantum dynamics.
The generalized quantum dynamics (GQD) developed in this way has been
shown [24,25] to open new possibilities for describing hadron dynamics.

In the present paper the GQD is used for
investigating quark-gluon retardation effects on low-energy hadron 
dynamics. We show that retardation from quark confinement results in an anomalous
off-shell behavior of the two-nucleon amplitudes and in a lack of continuity 
of the evolution operator describing low-energy hadron dynamics that
for this reason is not governed by the Schr{\"o}dinger equation.  In
Sec.II we discuss some problems in describing hadron dynamics that 
arise due  to quark-gluon retardation effects.
The principal features of the GQD are reviewed in Sec.III, and in Sec.IV we
show that the GQD provides an extension of Hamiltonian dynamics which can
describe the evolution of quantum systems with confined
degrees of freedom. The existence of such degrees of freedom gives rise to
a peculiar dynamical situation that allows one to conclude that retardation
effects of quark confinement on low-energy hadron can be significant.
In
Sec.IV this fact is illustrated by the example of the dynamics 
of nucleons with internal structure described by a
constituent quark model.
In Sec.VI we use the GQD to construct a model that is 
a generalization of the separable-potential model to the case when the $NN$
interaction is nonlocal in both space and time. We show that despite
its simplicity the model, in which retardation effects of quark
confinement are taken into account,
yields the nucleon-nucleon phase shifts in good agreement with experiment.
Analyzing the off-shell behavior of the two-nucleon amplitudes in the case
where the $NN$ interaction is nonlocal in time, we find that 
quark-gluon retardation 
can have significant effects on three- and many-nucleon results.

\section*{II. Retardation effects and low-energy hadron dynamics}

According to QCD, the physics of the strong interaction exhibits different
relevant excitations at distinct length (or momentum) scales. At short
space-time distances ($r<<1fm$) the relevant degrees of freedom are quarks
and gluons, effectively unconfined due to asymptotic freedom. At large
distances ($r>>1fm$), on the other hand, hadronic degrees of freedom
should be considered as relevant degrees of freedom. At the same time,
due to confinement there are no observables, which can be associated
with the quark and gluon degrees of freedom at least in the low-energy
regime. However, in principle this does not mean that the quark and
gluon degrees of freedom are not relevant in this regime. The above
means only that there are no observables associated
with quarks and gluons in the low-energy regime. To clarify
this point, note the following.   
The most fundamental aspect of quantum
theory in its present interpretation is its probabilistic character. The
description of a physical system requires two kinds of elements, the
observables and the states of the system. The states in turn are described
by the vectors of a Hilbert space. The state vector determines the
probabilities of finding some values of the observables, when a
measurement is performed, and can be expanded in terms of the eigenvectors
corresponding to the eigenvalues of a complete system of commuting
observables. From this and the phenomenon of quark confinement,
according to which quarks and gluons are not observable in free states,
it follows that vectors
describing low-energy states of a system of strongly interacting
particles can be expanded in terms of
eigenvectors associated with hadronic degrees of freedom, i.e. they
belong to the Hilbert space of hadron states. The time evolution
of the system is described by the evolution equation
\begin{equation}
|\psi (t)> = U(t,t_0)|\psi(t_0)>,
\end{equation}
where $|\psi (t)>$ is a state vector of the Hilbert space of hadron
states, and $U(t,t_0)$ is the evolution operator defined on this space.
Since due to confinement quark-gluon retardation effects must not
lead to a loss of probability from the Hilbert space,  $U(t,t_0)$
must be a unitary operator
\begin{equation}
U^{+}(t,t_0) U(t,t_0) = U(t,t_0) U^{+}(t,t_0) = {\bf 1}.
\end{equation}
Here we use the interaction picture. Thus due to confinement the
time evolution of a system of strongly interacting particles at low energies 
can be described by the unitary evolution operator defined on the 
Hilbert space of hadron states. Nevertheless, the quark and gluon
degrees of freedom may be relevant in describing low-energy hadron
dynamics, since they should manifest themselves in hadronic interactions.

Let us now show that quarks and gluons confined within hadrons 
can have substantial effects on low-energy hadron dynamics.
One of the fundamental requirement of quantum theory is that the evolution
operator must satisfy the composition law
\begin{equation}
U(t,t') U(t',t_0) = U(t,t_0), \quad U(t_0,t_0)= {\bf 1}.
\end{equation}
In the case of an isolated system, the evolution operator in the
Schr{\"o}dinger picture
$U_s(t_2,t_1) \equiv exp(-iH_0t_2) U(t_2,t_1) 
exp(iH_0t_1)$ depends on the difference $(t_2-t_1)$ only, so that 
the operators
$V(t) \equiv U_s(t,0)$  constitute a one-parameter group
of unitary operators, with the group property 
\begin{equation}
V(t_1+t_2) = V(t_1) V(t_2) , \quad    V(0)= {\bf 1}.
\end{equation}
Here and below, we use the units in which $c=\hbar =1,$
and $H_0$ is the free Hamiltonian. If the evolution operator
is assumed to be strongly continuous, i.e. if
\begin{equation}
\lim \limits_{t_2 \tend t_1} \Vert V(t_2)\vert\psi> - V(t_1) \vert \psi> 
 \Vert = 0,
\end{equation}
then from Stone's  theorem it follows that this one-parameter 
group has a self-adjoint infinitesimal  generator $H$: 
$$ V(t) = exp(-iHt), \quad   
 i d/dt	V(t)=H V(t).$$
Identifying $H$ with the total Hamiltonian as usual, we get 
the time-dependent  Schr{\"o}dinger equation: 
$ i \frac{d |\psi_s(t)>}{dt} = H |\psi_s(t)>,$ 
where $ |\psi_s(t)> = V(t)|\psi_s(t=0)>$ is a state vector in the
Schr{\"o}dinger picture. Thus, in this case, the interaction being
described by the interaction Hamiltonian is necessarily instantaneous.
Obviously, the infinitesimal generator $H$ being the operator of the
total energy cannot contain terms depending on energy i.e. on its spectral
parameter. From this it follows that, if the continuity condition (5) is
satisfied, then the time evolution of a system can be unitary only in the
case when the interaction generating the dynamics of a quantum system is
instantaneous. Thus the evolution operator describing low energy hadron
dynamics cannot be strongly continuous, since due to quark-gluon
retardation effects hadron interactions should be nonlocal in time,
and, on the other hand, due to confinement the evolution operator
must be unitary. This means that quark-gluon retardation and confinement 
must result in the fact that the
evolution of hadron systems cannot be governed by the Schr{\"o}dinger
equation. 

We have shown that quark-gluon retardation effects give rise to the
fact that the evolution operator describing hadron 
dynamics cannot be strongly continuous. On the other hand, although 
in the Hamiltonian formalism
the requirement that the evolution operator 
must be strongly continuous is used
as a fundamental assumption (due to Stone's theorem it is equivalent to the
assumption that the evolution of a quantum system is governed by the
Schr{\"o}dinger equation), this requirement is not necessary on physical
grounds, and it is enough to require that
\begin{equation}
 <\psi_2| V(t_2)|\psi_1>\tend\limits_{t_2 \tend t_1} <\psi_2|V(t_1)|\psi_1> 
\end{equation}
for any physically realizable states $|\psi_1>$ and $|\psi_2>$ [26].
Note, in this connection, that there are normalized vectors in the Hilbert
space that represent the states for which the energy of a system is
infinite. Such states cannot be considered as physically realizable,
and hence the corresponding matrix elements of the evolution operator need
not be continuous. From this it follows that in principle Hamiltonian 
dynamics can be extended to the case where only the physical continuity 
condition (6) is satisfied. As has been shown in Ref.[21], this
really can be done by using the Feynman approach to quantum theory in
combination with the canonical approach. As we
show bellow, the GQD developed in this way allows one to take into account
quark-gluon retardation in describing hadron dynamics.

\section*{III. Generalized Quantum Dynamics}

In the GQD the following assumptions are used as basic
postulates: 

(i) The physical state of a system is represented by a vector
(properly by a ray) of a Hilbert space.

(ii) An observable A is represented by a Hermitian hypermaximal operator
$\alpha$. The eigenvalues $a_r$ of $\alpha$ give the possible values of A.
An eigenvector $|\varphi_r^{(s)}>$ corresponding to the eigenvalue
$a_r$ represents a state in which A has the value $a_r$. If the system
is in the state $|\psi>,$ the probability $P_r$ of finding the value
$a_r$ for A, when a measurement is performed, is given by
$$P_{r} = <\psi|P_{V_{r}} |\psi>= \sum_s |<\varphi_r^{(s)}|\psi>|^2, $$
where $P_{V_{r}}$ is the projection operator on the eigenmanifold $V_r$
corresponding to $a_r,$ and the sum $\Sigma_s$ is taken over a complete
orthonormal set ${|\varphi_r^{(s)}>}$ (s=1,2,...) of $V_r.$
The state of the system immediately after the observation is
described by the vector $P_{V_{r}}|\psi>.$

These  assumptions are the main assumptions on which quantum
theory is founded.
In the canonical formalism these postulates 
are used together with the assumption that the time 
evolution of a state vector is governed by the
Schr{\"o}dinger equation.
In the formalism [21] this assumption is not used. 
Instead the assumptions (i) and (ii) are used together
with the following postulate.

(iii) The probability of an event is the absolute square of a 
complex number called the probability amplitude. 
The joint probability amplitude of a time-ordered 
sequence of events is product of the separate probability 
amplitudes of each of these events. 
The probability amplitude of an event which can happen in several 
different ways is a sum of the probability 
amplitudes for each of these ways.

The statements of the assumption (iii) express the well known 
law for the quantum-mechanical probabilities. 
Within the canonical formalism this law is derived 
as one of the consequences of the theory. 
However, in the Feynman formulation of 
quantum theory this law is directly derived starting from 
the analysis of the phenomenon of quantum interference, 
and is used as a basic  postulate of the  theory.
According to the assumption (iii), the probability amplitude of an 
event which can happen in several different ways is a 
sum of contributions from each alternative way. 
In particular, the amplitude 
 $<\psi_2| U(t,t_0)|\psi_1>$ can be represented as a sum 
of contributions from all alternative
ways of realization of the corresponding evolution process.
Dividing these alternatives in different classes, we can then analyze
such a probability amplitude in different ways.
For example, subprocesses with definite instants of the
beginning and  end of the interaction in the system
can be considered as such alternatives.
In this way the amplitude
$<\psi_2|U(t,t_0)|\psi_1>$  can be written in the form [21]
\begin{equation}
<\psi_2| U(t,t_0)|\psi_1> = <\psi_2|\psi_1> + 
\int_{t_0}^t dt_2 \int_{t_0}^{t_2} dt_1
<\psi_2|\tilde S(t_2,t_1)|\psi_1>,
\end{equation}
where 
$<\psi_2|\tilde S(t_2,t_1)|\psi_1>$ is the probability amplitude 
that if at time $t_1$ the system was in the state $|\psi_1>,$ then the 
interaction in the system will begin at time $t_1$ and 
will end at  time $t_2,$ and at this time the 
system will be in the state $|\psi_2>.$ 
More precisely,  $<\psi_2|\tilde S(t_2,t_1)|\psi_1>$ is the density of the
probability amplitude of the above event, and as contributions to
$<\psi_2|U(t,t_0)|\psi_1>$ one has to consider amplitudes 
$<\psi_2|\tilde U_\varepsilon(t_2,t_1)|\psi_1>=\int_{t_2}^{t_2+\varepsilon}
dt'_2\int_{t_1}^{t_1+\varepsilon}dt'_1\theta(t'_2-t'_1)
<\psi_2|\tilde S(t'_2,t'_1)|\psi_1>,$ where $\varepsilon<<t_2-t_1$ and
$$
\theta(\tau)=\left \{
\begin{array}{rcl}
1,\quad for \quad \tau> 0\\
0,\quad for \quad \tau<0.\\
\end{array}
\right.
$$
The amplitude $<\psi_2|\tilde U_\varepsilon(t_2,t_1)|\psi_1>$ is the 
amplitude that if at time $t_1$ the system was in the state $|\psi_1>$,
then the interaction in the system will begin in time internal 
$t_1,$ $t_1+\varepsilon$ and will end in time interval $t_2,$ $t_2+
\varepsilon$, and at the time $t_2+\varepsilon$ the system will in the 
state $|\psi_2>$. In general $\tilde S(t_2,t_1)$ may be only an 
operator-valued 
generalized function of $t_1$ and $t_2$ [21]. Nevertheless, it is
convenient to call $\tilde S(t_2,t_1)$ an "operator", using this word
in generalized sense.
In the case of an isolated system the operator
$\tilde S(t_2,t_1)$ can be represented in the form [21]
\begin{equation}
\tilde S(t_2,t_1) = exp(iH_0t_2)
\tilde T(t_2-t_1) exp(-iH_0 t_1).
\end{equation}

To clarify the role, which  the operator $\tilde S(t_2,t_1)$ 
plays in the GQD, note the following. 
The  Feynman formulation is based on the assumption 
that the history of a system can be  represented by some path in 
space-time. 
From the postulate (iii) it then follows that the 
probability amplitudes of any event  is 
a sum of the probability amplitudes that a particle 
has a completely specified path in space-time. 
The contribution from a single path is postulated to be an 
exponential whose (imaginary) phase is the classical action 
(in units of $\hbar$) for the path in question. 
In the GQD the history of a system is 
represented by the version  of the time evolution 
of the system associated with completely specified instants of 
the beginning and end of the interaction in the system.
Such a description of the history of a system is  
more general and requires no supplementary 
postulates like the above assumptions of the  Feynman formulation.
On the other hand, the probability  amplitudes 
$<\psi_2|\tilde S(t_2,t_1)|\psi_1>$, in terms of which we describe 
quantum dynamics, are used in the spirit of Feynman's theory: 
The probability amplitude of any event is  
represented as a sum of these amplitudes. 

As has been shown in Ref.[21], for the evolution operator $U(t,t_0)$ given
by (7) to be unitary
for any times $t_0$ and $t$, 
the operator $\tilde S(t_2,t_1)$ must satisfy the following
equation:
\begin{equation}
(t_2-t_1) \tilde S(t_2,t_1) = 
\int^{t_2}_{t_1} dt_4 \int^{t_4}_{t_1}dt_3
(t_4-t_3) \tilde S(t_2,t_4) \tilde S(t_3,t_1).
\end{equation}
Note that, since $\tilde S(t_2,t_1)$ may be only an operator-valued
distribution, in general the production $\tilde S(t_2,t_4)\tilde S(t_3,t_1)$
is not defined at $t_4=t_3$. However, this does not lead to any problems
because in Eq.(9) the above production is multiplied by the factor
$t_4-t_3$.
Eq.(9) allows one to obtain the operators $\tilde S(t_2,t_1)$
for any $t_1$ and $t_2$, if the operators
$\tilde S(t'_2, t'_1)$ corresponding to infinitesimal 
duration times $\tau = t'_2 -t'_1$ of interaction are known.
Thus this equation allows one to obtain the contributions to the 
evolution operator
from the processes associated with any duration times of interaction
if those from the processes associated with infinitesimal duration times of
interactions are known. It is natural to assume that most of the 
contribution to the evolution operator 
in the limit $t_2 \to t_1$ comes from the processes 
associated with an fundamental interaction  
in the system under study.
Denoting this contribution by 
$H_{int}(t_2,t_1)$, the operator 
$\tilde S(t_2,t_1)$ can be represented in the form 
\begin{equation}
\tilde{S}(t_2,t_1) = 
H_{int}(t_2,t_1) + \tilde{S}_1(t_2,t_1), 
\end{equation}
where $\tilde{S}_1(t_2,t_1)$ is the part of the 
operator $\tilde{S}(t_2,t_1)$ which in the limit 
$t_2 \tend t_1$ gives a 
negligibly small contribution to the evolution 
operator in comparison with $H_{int}(t_2,t_1).$
Within the GQD 
the operator $H_{int}(t_2,t_1)$ 
plays the role which the interaction 
Hamiltonian plays in the ordinary formulation of quantum theory:
It generates dynamics of a system. 
This operator can be regarded as a generalization of the 
interaction Hamiltonian, and it was called the
generalized interaction operator. Obviously, the operator  $H_{int}(t_2,t_1)$
must satisfy Eq.(9) in the limit $t_2 \to t_1$ 
\begin{equation}
F(t_2,t_1) \tend \limits_{t_2 \to t_1} 0,
\end{equation}
where
$$
F(t_2,t_1) = 
- (t_2-t_1) H_{int}(t_2,t_1) +
\int^{t_2}_{t_1} dt_4 \int^{t_4}_{t_1} dt_3 
(t_4-t_3) H_{int}(t_2,t_4) H_{int}(t_3,t_1).
$$
If  $H_{int}(t_2,t_1)$ is specified, Eq.(9) allows one to find the
operator $\tilde S(t_2,t_1).$ 
Formula (7) can then be used to construct the evolution operator 
$U(t,t_0)$ and accordingly the state vector 
\begin{equation}
|\psi(t)> = |\psi(t_0)> +  \int_{t_0}^t dt_2
\int_{t_0}^{t_2} dt_1 \tilde S(t_2,t_1) |\psi(t_0)> 
\end{equation}
at any time $t.$ The corresponding evolution operator is of the form [21]
\begin{equation}
U(t,t_0)=  {\bf 1}+ \frac{1}{2\pi}  
\int^\infty_{-\infty} dx 
 \frac {exp[-i(z-H_0)t] T(z) exp[i(z-H_0)t_0]}
{(z-H_0)(z-H_0)},  
\end{equation}
where $z=x+iy$, $x$ and $y$ are real, and $y>0$, 
and the operator $T(z)$ is defined by
\begin{equation}
T(z) = i \int_{0}^{\infty} d\tau exp(iz\tau)
\tilde T(\tau),
\end{equation}
with $\tilde T(\tau) = exp(-iH_0t_2) \tilde S(t_2,t_1)
exp(iH_0t_1)$.
Thus Eq.(9) can be regarded as an equation of motion for states 
of a quantum system. 
It should be noted that in the GQD the continuity
condition (6) is not used as a basic postulate, and plays only the role
of a consistency condition of the theory: Matrix elements of the evolution
operator (7) obtained by solving Eq.(9) must be continuous for any
physically realizable states. It is reasonable to consider such states as 
the states for which $\Vert H_0|\psi>\Vert<\infty$. Thus the condition
(6) must be satisfied for all vector $|\psi>\in {\cal D}(H_0)$, 
${\cal D}(H_0)$ being domain of $H_0$.

From the  mathematical point of view the 
requirement that $H_{int}(t_2,t_1)$ contains all the 
dynamic information that is needed for constructing 
$U(t_2,t_1)$ means that the operator $H_{int}(t_2,t_1)$ 
must have such a form that Eq.(9) 
has a unique solution having the following behavior 
near the point $t_2=t_1:$
\begin{equation}
\tilde S(t_2,t_1)
\tend \limits_{t_2 \tend t_1}
 H_{int}(t_2,t_1) 
+ o(\tau^{\epsilon}),
\end{equation}
where $\tau=t_2-t_1$ and the value of $\epsilon$ 
depends on the form of the operator $ H_{int}(t_2,t_1).$
Since $\tilde S(t_2,t_1)$ and $ H_{int}(t_2,t_1)$ are only 
operator-valued distributions, the mathematical meaning of the condition (15)
needs to be clarified. We will assume that the condition (15) means 
that
$$
<\psi_2|\left (\int_{t_0}^{t}
dt_2\int_{t_0}^{t_2}dt_1\tilde S(t_2,t_1)\right)|\psi_1>\tend\limits_{t\tend
t_0}
$$
\begin{equation}
<\psi_2|\left (\int_{t_0}^{t}dt_2\int_{t_0}^{t_2}dt_1H_{int}
(t_2,t_1)\right)|\psi_1>+o(\tau^{\alpha+2}),
\quad \tau=t-t_0,
\end{equation}
for any vectors $|\psi_1>$ and $|\psi_2>$  of the Hilbert space. Note
that the condition (11) has to be considered in the same sense.

To clarify the above statement,  
note that Eq.(9) is equivalent to 
the following equation [21]:
\begin{equation}
\frac{d T(z)}{dz} =
 T(z) G^{(2)}(z) T(z) ,
\end{equation}
with
$$
 G^{(2)}(z) = - \sum \limits_{n} 
\frac{|n><n|}{(z-E_n)^2}.
$$
Here  $n$ stands for the entire set of discrete and continuous
variables that characterize the system in full, and
$|n>$ are the eigenvectors of the free Hamiltonian:
$H_0|n>= E_n|n>.$ Thus, instead of solving
Eq.(9), one can solve Eq.(17) for the operator $T(z)$. The operator
$\tilde T(\tau)$ 
and correspondingly the operator $\tilde S(t_2,t_1)$ can then be obtained
by using the Fourier transform
\begin{equation}
<n_2|\tilde T(\tau)|n_1> = - \frac{i}{2\pi} \int _{-\infty}^{\infty}dx
exp(-iz\tau) <n_2|T(z)|n_1>,
\end{equation}
where $z=x+iy,$ $x$ and $y$ are real, and $y>0.$ At the same time,
the operator $T(z)$ can be directly used for constructing 
the evolution operator. 
According to (14) and (15), the operator $T(z)$ 
has the following asymptotic behavior 
for $|z| \tend \infty:$ 
\begin{equation}
<n_2|T(z)|n_1> \tend \limits_{|z| \tend \infty}
<n_2| B(z)|n_1> + o(|z|^{-\beta}) ,
\end{equation}
where
$$
B(z) = i \int_0^{\infty} d\tau exp(iz \tau)
H^{(s)}_{int}(\tau).
$$
$\beta=1+\epsilon,$ and
$H^{(s)}_{int}(t_2-t_1) = exp(-iH_0t_2) H_{int}(t_2,t_1)
exp(iH_0t_1)$ is the generalized 
interaction operator in the Schr{\"o}dinger picture. 
From (17) and (19) it follows that the operator $B(z)$ must 
satisfy the following asymptotic condition:
$$
\frac{d <n_2|B(z)|n_1>}{dz} 
\tend \limits_{|z| \tend \infty}
<n_2| B(z) G^{(2)}(z) B(z)|n_1>+ o(|z|^{-\beta}).
$$
The above requirements, which the generalized interaction operator 
has to meet, mean that 
$B(z)$ must be so close to the solution of 
equation (17) in the limit $|z| \tend \infty$ that this 
differential equation has a unique solution 
having the asymptotic behavior (19). 
The operator $B(z)$ represents the contribution 
which $H^{(s)}_{int}(\tau)$ gives to the 
operator $T(z),$ and was called the 
effective interaction operator. It should be also noted,
that the operator $T(z)$ satisfies the equation
\begin{equation}
T(z_1) - T(z_2) = (z_2 -z_1) \sum_n 
 \frac {T(z_2)|n><n|T(z_1)}
{(z_2-E_n)(z_1-E_n)},
\end{equation}
provided that $\tilde S(t_2,t_1)$ satisfies Eq.(9).

As has been shown in Ref.[21], the dynamics governed by Eq.(9)
is equivalent to  Hamiltonian dynamics in the case where the
generalized interaction operator is of the form 
\begin{equation}
 H_{int}(t_2,t_1) = - 2i \delta(t_2-t_1) 
 H_{I}(t_1) ,
\end{equation}
$H_{I}(t_1)$ being the interaction Hamiltonian in the interaction
picture. In this case the state vector $|\psi(t)>$ 
given by (12) satisfies the 
Schr{\"o}dinger equation
$$
  \frac {d |\psi(t)>}{d t} = -iH_I(t)|\psi(t)>.
$$
The delta function $\delta(\tau)$ in (21) 
emphasizes that in this case 
the fundamental interaction is instantaneous. 
Thus the Schr{\"o}dinger equation results from the generalized equation of
motion (9) in the case where the interaction
generating the dynamics of a quantum system is 
instantaneous. At the same time,
Eq.(9) permits the generalization to the case where the operator
$H_{int}(t_2,t_1)$ has no such a singularity as the delta function
at the point $t_2=t_1$ [21].
In this case the fundamental interaction 
generating the dynamics of a quantum system is 
nonlocal in time:
The evolution operator is defined by 
$H_{int}(t_2,t_1)$ as a function of the time 
duration $\tau=t_2-t_1$ of the interaction. In a more 
general case, the generalized interaction operator has the 
following form [25]:
\begin{equation}
H_{int}(t_2,t_1)=-2i\delta(t_2-t_1)H_I(t_1)+H_{non}(t_2,t_1),
\end{equation}
where the first term on the right-hand side of (22) describes the
instantaneous part of the interaction generating the dynamics 
of a quantum system, while the term $H_{non}$
represents its nonlocal-in-time part. In order that the 
dynamics governed by Eq.(9) be different from the 
dynamics governed by the Schr{\"o}dinger equation with the 
interaction Hamiltonian $H_I(t)$, the nonlocal operator $H_{non}$ 
must satisfy the condition 
\begin{equation}
(t_2-t_1) H_{non}(t_2,t_1) \sim
\int^{t_2}_{t_1} dt_4 \int^{t_4}_{t_1} dt_3 
(t_4-t_3) H_{non}(t_2,t_4) H_{non}(t_3,t_1), \quad t_2\tend t_1.
\end{equation}
In fact, in the local case, where $H_{int}(t_2,t_1)$ is of the
form (21), the first term in the operator $F(t_2,t_1)$
is zero, and from the condition (11) it follows that the second
term in this operator must tend to zero in the limit
 $t_2 \to t_1$ so rapidly as it is required for Eq.(9) to have
a unique solution. In the nonlocal case, both terms in the
operator $F(t_2,t_1)$ are not zero, and, as we show in Sec.IV, they
do not decrease when $t_2 \to t_1$ so rapidly as it is
needed. Only their sum can vanish fast enough for the condition (11)
to be satisfied. From
this it follows that the nonlocal part of the generalized
interaction operator must satisfy the condition (23).

The principal feature of the GQD is that it provides an extension of
Hamiltonian dynamics which can describe the unitary evolution of a quantum
system with confined degrees of freedom.  In the next
section we will show that the existence of such degrees of freedom,
that can manifest themselves only
in some retardation of the interaction in the system, gives rise
to a peculiar dynamical situation that allows one to conclude that
retardation effects of quark confinement on low-energy 
hadron dynamics can be very significant.

\section*{IV.Effects of confined degrees of freedom on the dynamics of a 
quantum system}

Let us consider the evolution problem for two nonrelativistic particles 
in the c.m.s. 
We denote the relative momentum by ${\bf p}$ and 
the reduced mass by $\mu.$ 
Assume that the generalized interaction operator in the 
Schr{\"o}dinger picture $H^{(s)}_{int}(\tau)$ 
has the form 
\begin{equation}
<{\bf p}_2| H^{(s)}_{int}(\tau) |{\bf p}_1> = 
\varphi^*({\bf p}_2) \varphi({\bf p}_1) f(\tau), 
\end{equation}
where $f(\tau)$ is some function of $\tau,$ and 
the form factor $\varphi ({\bf p})$ has the 
following asymptotic behavior for $|{\bf p}| \tend \infty:$
\begin{equation}
\varphi({\bf p}) \sim \frac {c_1} {|{\bf p}|^{\alpha}}, 
\quad  {(|{\bf p}| \tend \infty).}
\end{equation}
Let, for example, $\varphi ({\bf p})$ be of the form
\begin{equation}
\varphi({\bf p}) = 
\frac {c_1} {|{\bf p}|^{\alpha}}+g({\bf {p}}),
\end{equation}
and in the limit $|{\bf p}| \tend \infty$ the function $g({\bf {p}})$ 
satisfies the estimate $g({\bf {p}})=o(|{\bf {p}}|^{-\delta})$, where
$\delta>\alpha,$ 
$\delta>\frac{3}{2}.$
In this case, the problem can be easily solved by 
using Eq.(15). 
Representing $<{\bf p}_2| T(z) |{\bf p}_1>$ in the form 
$
<{\bf p}_2| T(z)|{\bf p}_1> = 
\varphi ({\bf p}_2)\varphi^* ({\bf p}_1) t(z),
$
from (15), we get the equation 
\begin{equation}
\frac {dt(z)}{dz} = 
-t^2(z) \int d^3k \frac {|\varphi ({\bf k})|^2}
{(z-E_k)^2}
\end{equation}
with the asymptotic condition 
\begin{equation}
t(z)  \tend \limits_{|z| \tend -\infty} 
f_1(z) + o(|z|^{-\beta}),
\end{equation}
$f_1(z)= i \int_0^{\infty} d\tau 
exp(iz\tau) f(\tau), $
and $E_k = \frac {k^2}{2 \mu}.$ 
The solution of Eq.(27) 
with the initial condition $t(a)=g_a,$ where $a \in (-\infty,0),$ is 
\begin{equation}
t(z) = g_a \left(1 +(z-a) g_a 
 \int d^3k \frac {|\varphi ({\bf k})|^2}
{(z-E_k)(a-E_k)} \right)^{-1}.
\end{equation}
In the case $\alpha >\frac{1}{2}$, the 
function $t(z)$ tends to a constant as 
$z \tend -\infty$:
\begin{equation}
t(z)  \tend \limits_{z \tend -\infty}  
\lambda.
\end{equation}
Thus in this case the function $f_1(z)$ must 
tend to $\lambda$ as $z \tend -\infty.$ 
From this it follows that the only 
possible form of the function $f(\tau)$ is 
$$
f(\tau) = -2i \lambda \delta(\tau) 
+ f^{\prime}(\tau),
$$
where the function $f^{\prime}(\tau)$ 
has no such a singularity at the point $\tau=0$ 
as the delta function.
In this case  the generalized interaction operator 
$H^{(s)}_{int}(\tau)$ has the form (21) and hence the 
dynamics generated by this operator is equivalent 
to the dynamics governed by the 
Schr{\"o}dinger equation with the separable potential
\begin{equation}
<{\bf p}_2|H_I|{\bf p}_1> = \lambda \varphi^*({\bf p}_2) 
\varphi({\bf p}_1).
\end{equation}
Solving Eq.(27) with the boundary condition (30), 
we easily get the well-known expression for the 
T matrix in the separable-potential model
\begin{equation}
<{\bf p}_2|T(z)|{\bf p}_1> = 
\lambda \varphi^* ({\bf p}_2)\varphi({\bf p}_1) 
\left (1 - \lambda \int d^3k \frac 
{|\varphi({\bf k})|^2}{z-E_k} \right )^{-1}.
\end{equation}

Ordinary quantum mechanics does not permit the 
extension of the above model to the case 
$\alpha \leq \frac{1}{2}.$
Indeed, in the case of such a large-momentum behavior of 
the form factors $\varphi({\bf p}),$ 
the use of the interaction Hamiltonian given by (31) 
leads to the ultraviolet divergences, i.e.
the integral in (32) is not convergent. 
We will now show that our formalism allows one to 
extend this model to the case $0 < \alpha <\frac{1}{2}.$
Let us determine the class of the functions $f_1(z)$ 
and correspondingly the value of $\beta$ for 
which Eq.(27) has a unique solution having the 
asymptotic behavior (28).
In the case $\alpha <\frac{1}{2},$ the 
function $t(z)$ given by (29) has the following 
behavior for $z \tend -\infty:$
\begin{equation}
t(z)  \tend \limits_{z \tend -\infty}  b_1 (-z)^{\alpha-\frac{1}{2}}+
b_2 (-z)^{2 \alpha-1} + o(|z|^{2 \alpha-1}),
\end{equation}
where $b_1 =- \frac{1}{2} cos(\alpha \pi) \pi^{-2} |c_1|^{-2} 
(2 \mu)^{\alpha-\frac{3}{2}}$ and $b_2= b_1 |a|^{\frac{1}{2}-
\alpha} -b_1^2(M(a)+g_a^{-1})$ with
$$
M(a) = \int \frac {|\varphi({\bf k})|^2-\frac {|c_1|^2}
 {|{\bf {k}}|^{2\alpha}}}
{a-E_k}d^3k .
$$
The parameter $b_1$ does not depend on $g_a.$ This means that all 
solutions of Eq.(27) have the same 
leading term  in (33), and only the second term distinguishes 
the different solutions of this equation. 
Thus, in order to obtain a unique solution of Eq.(27), 
we must specify the first two terms in the 
asymptotic behavior of $t(z)$ for $z \tend - \infty.$ 
From this it follows that the functions 
$f_1(z)$ must be of the form 
\begin{equation}
f_1(z) = b_1 (-z)^{\alpha-\frac{1}{2}} +
b_2 (-z)^{2 \alpha -1} ,
\end{equation}
and $\beta=2 \alpha -1.$ 
Correspondingly the functions $f(\tau)$ must be of the form
\begin{equation}
f(\tau) = a_1 \tau^{-\alpha-\frac{1}{2}} +
a_2 \tau^{-2 \alpha} ,
\end{equation}
with $a_1= -ib_1 \Gamma ^{-1}(1-2\alpha) exp[i(-\frac{\alpha}{2}+
\frac{1}{4}) \pi],$ and 
$a_2= -b_2 \Gamma ^{-1}(1-2\alpha) exp(-i \alpha \pi),$
where $\Gamma(z)$ is the gamma-function.
This means that in the case 
$\alpha <\frac{1}{2}$ the generalized interaction operator 
must be of the form 
$$
<{\bf p}_2| {H}^{(s)}_{int}(\tau)|{\bf p}_1> = f(\tau)\varphi^* ({\bf
p}_2)\varphi ({\bf p}_1)=
$$
\begin{equation}
=a_1 \varphi^* ({\bf p}_2)\varphi^* ({\bf p}_1) \tau^{-\alpha-\frac{1}{2}} +
a_2 \varphi^* ({\bf p}_2)\varphi ({\bf p}_1) \tau^{-2 \alpha},
\end{equation}
and, as it follows from (29) and (33), for the T matrix  we have
\begin{equation}
<{\bf p}_2| T(z)|{\bf p}_1> = 
N(z) \varphi^* ({\bf p}_2)\varphi ({\bf p}_1), 
\end{equation}
with
$$
N(z) = g_a \left (1 + (z-a) g_a \int d^3 k
\frac{|\varphi({\bf k})|^2}
{(z-E_k)(a- E_k)} \right )^{-1},
$$
where  
$
g_a = b_1^2\left (b_1 |a|^{\frac{1}{2} -\alpha} + a_2 \Gamma(1-2\alpha)
exp( i \alpha \pi)-b_1^2M(a)\right )^{-1}. 
$
It can be easily checked that 
$N(z)$ can be represented in the following form
$$
N(z)=\frac{b_1^2}{-b_2+b_1(-z)^{\frac{1}{2}-\alpha}+M(z)b_1^2}.
$$

By using (13) and (37), we can construct the 
evolution operator 
$$
<{\bf p}_2|U(t,t_0)|{\bf p}_1> = <{\bf p}_2|{\bf p}_1>
 - \frac {i}{2\pi} 
\int_{-\infty}^{\infty} dx
$$
\begin{equation}
\times 
\frac {exp[-i(z-E_{p_2})t] exp[i (z - E_{p_1})t_0]}
{(z-E_{p_2})(z-E_{p_1})} 
N(z) \varphi^*({\bf p}_2) \varphi({\bf p}_1), 
\end{equation}
where  $z=x+iy,$ and $y>0.$ 
The evolution operator $U(t,t_0)$ defined by (38) is a unitary 
operator satisfying the composition law (3), provided that the 
parameter $b_2$ is real.

We have stated the correspondence between the form 
of the generalized interaction operator and 
the large-momentum behavior of the form 
factor $\psi({\bf p}).$ 
In the case $\alpha >\frac{1}{2},$ the operator 
$H^{(s)}_{int}(\tau)$ would necessarily have 
the form (21). 
In this case the fundamental interaction is 
instantaneous. 
In the case $0 <\alpha <\frac{1}{2}$ (the restriction $\alpha >0$ 
is necessary for the integral in (29) to be convergent), 
the only possible form of 
$H^{(s)}_{int}(\tau)$ is (36), and hence the interaction 
generating  the dynamics of the system is 
nonlocal in time.
Thus the interaction generating the dynamics can be nonlocal in time
only if the form factors have the "bad" large-momentum behavior that
within Hamiltonian dynamics gives rise to the ultraviolet divergences.

Let us now show that in the case $\alpha<\frac{1}{2}$ the dynamics
generated by the generalized interaction operator (36) is not 
equivalent to Hamiltonian 
dynamics. The evolution operator $U(t,t_0)$ given by (38) satisfies the
composition law (3) for any $t_0$, $t$ and $t'$, provided
the operator $\tilde S(t_2,t_1)$ satisfies Eq.(9) for any $t_1$ and $t_2$.
In the case of model under study the operator $\tilde S(t_2,t_1)$ is of the 
form
\begin{equation}
\tilde S(t_2,t_1)=D(t_2,t_1)\tilde F(t_2,t_1),
\end{equation}
where $\tilde F(\tau)$ is a function of $\tau$, and $D(t_2,t_1)=exp(iH_0t_2)
|\varphi><\varphi|exp(-iH_0t_1)$ is the operator-valued distribution
such that
\begin{equation}
<{\bf p_2}|D(t_2,t_1)|{\bf p_1}>=exp(iE_{p_2}t_2)
<{\bf p_2}|\varphi><\varphi|{\bf p_1}>exp(-iE_{p_1}t_1),
\end{equation}
with $<\varphi|{\bf p}>=\varphi({\bf p}).$ By using (39) and (40), 
for the function
 $\tilde F(\tau)$ we get the following equation:
$$
(t_2-t_1) \tilde F(t_2-t_1) = 
\int^{t_2}_{t_1} dt_4  \int_{t_1}^{t_4} dt_3 (t_4-t_3)\int d^3k
exp[-iE_k(t_4-t_3)]\times
$$
\begin{equation}
\times <\varphi|{\bf k}><{\bf k}|\varphi>
\tilde F(t_2-t_4)\tilde F(t_3-t_1).
\end{equation}
Let us examine this equation in the limit $t_2\tend t_1$. For this we have
to change the variables: $\tau_i=\theta_i/\nu,$ ${\bf k}=\nu{\bf q_\nu},$
$i=1,2,3,4$. With such a change of the variables, Eq.(41) can be rewritten 
in the form
$$
(\theta_2-\theta_1) \tilde F(\theta_2/\nu^2-\theta_1/\nu^2) 
= \nu^{-4}\int^{\theta_2}_{\theta_1} d\theta_4  \int_{\theta_1}^{\theta_4} 
d\theta_3 (\theta_4-\theta_3)\int d^3k
exp[-iE_{q_\nu}(\theta_4-\theta_3)]\times
$$
\begin{equation}
\times <\varphi|{\bf q}_\nu><{\bf q}_\nu|\varphi>
\tilde F(\theta_2/\nu^2-\theta_4/\nu^2)
\tilde F(\theta_3/\nu^2-\theta_1/\nu^2),
\end{equation}
$|{\bf q}_\nu>$ being the basis vectors such that $<{\bf k}|{\bf q}_\nu>=
\nu^{-\frac{3}{2}}\delta({\bf k}-\nu{\bf q}_\nu).$ The completeness
condition for these basis states reads  
\begin{equation}
\int d^3 q_\nu|{\bf q}_\nu><{\bf q}_\nu|={\bf 1}.
\end{equation}
Thus $|{\bf k}>$ and $|{\bf q}_\nu>$ are eigenstates of the operator
$H_0$ corresponding to distinct momentum scales. Since $<\varphi|{\bf q}_\nu>
= c_1|\nu{\bf q}_\nu|^{-\alpha}+o(\nu^{-\frac{3}{2}-\alpha})$
when $\nu\tend\infty$, by letting $t_2\tend t_1$,  from (42) we get
$$
(\theta_2-\theta_1) \tilde F(\theta_2/\nu^2-\theta_1/\nu^2) 
= \nu^{-1-2\alpha}\int^{\theta_2}_{\theta_1} d\theta_4  
\int_{\theta_1}^{\theta_4} d\theta_3 
\tilde F(\theta_2/\nu^2-\theta_4/\nu^2)\times
$$
\begin{equation}
\times\tilde F(\theta_3/\nu^2-\theta_1/\nu^2)
I(\theta_4-\theta_3,\nu),\quad \nu\tend\infty,
\end{equation}
with
$$I(\theta,\nu)=|c_1|^2\theta\int d^3q_\nu exp(-iE_{q_\nu}\theta)|{\bf
q}_\nu|^{-2\alpha}+o(\nu^{-\frac{3}{2}+\alpha}).$$
This expression emphasizes that the relevant momentum scale of intermediate
states
in Eq.(42) tends to infinity as $t_2\tend t_1$. In order to demonstrate that 
Eq.(44)
plays the key role in the case $\alpha<\frac{1}{2}$, note that from this
equation it follows that in the limit $\tau\tend 0$ the function
$\tilde F(\tau)$ behaves like $a_1\tau^{-\alpha-\frac{1}{2}}$
with
$$
a_1=(\theta_2-\theta_1)^{\frac{1}{2}-\alpha}\left(\int^{\theta_2}_{\theta_1} 
d\theta_4  \int_{\theta_1}^{\theta_4} d\theta_3I(\theta_4-\theta_3,\infty)
\right)^{-1}=
$$
\begin{equation}
=\frac{i}{2}\pi^{-2}|c_1|^{-2}(2\mu)^{\alpha-\frac{3}{2}}\Gamma^{-1}(1-2\alpha)
\cos(\alpha\pi)exp[i(-\frac{\alpha}{2}+\frac{1}{4})\pi],
\end{equation}
i.e. this
equation determines the main term of the generalized interaction operator
(see Eq.(36)). We see from (45) that only the intermediate states with
infinite momentum contribute to the parameter $a_1$. This manifests itself
in the fact that this parameter depends only on the leader term in the form
factor (25). Note in this connection that for
any vector $|\psi>$ of the Hilbert space ${\cal H}$ represented in the form
\begin{equation}
|\psi>=\int\psi({\bf k})|{\bf k}>d^3 k,
\end{equation}
with $\psi({\bf k})=<{\bf k}|\psi>$, one can construct another vector
\begin{equation}
|\psi_{\nu}>=\int\psi({\bf k}/\nu)\nu^{-\frac{3}{2}}|{\bf k}>d^3 k,
\end{equation}
that represents the same physical state, if we scale ${\bf k}\to\nu{\bf k}$,
i.e.
\begin{equation}
|\psi_{\nu}>=\int\psi({\bf q_\nu})|{\bf q_\nu}>d^3q_\nu,
\end{equation}
where ${\bf q_\nu}={\bf k}/\nu$.
Varying the parameter $\nu$, we get a set of vectors having 
the same norm
\begin{equation}
\Vert|\psi_{\nu}>\Vert=(\int\psi^*({\bf q_\nu})\psi({\bf q_\nu})d^3q_\nu)^
{\frac{1}{2}}.
\end{equation}
Each of the vectors $|\psi_{\nu}>$ 
belong to the Hilbert space ${\cal H}$ even when the parameter $\nu$ is letting
to infinity. To clarify this point, let us consider the state
$|\psi>=\int\psi_{k_0}({\bf k})|{\bf k}>d^3 k,$ where $\psi_{k_0}
({\bf k})$ is zero everywhere outside the subset $\Delta(k_0)$ $(E_{k_0}
\leq E_k\leq E_{k_0}+\varepsilon E_{k_0},$ $\varepsilon\ll 1)$ of the spectrum 
of $H_0$. Such a state is eigenstate of the projection operator $P_{\Delta
(k_0)}$ on the subset $\Delta(k_0)$. Correspondingly the state
\begin{equation}
|\psi_{\nu}>=\int\psi_{q_0}({\bf q_\nu})|{\bf q_\nu}>d^3q_\nu
\end{equation}
is eigenstate of the projection operator on the subset $\Delta(q_0)$
$(E_{q_0}\leq E_k\leq E_{q_0}+\varepsilon E_{q_0},\quad E_{q_0}=E_{k_0}
\nu^{-2})$. The projection operators are defined for any subsets even
when their location tend to the infinite part of the spectrum. From this
it follows that even when $\nu$ tends to infinity eigenvectors of the
projection operator $P_{\Delta(q_0)}$ belong to the Hilbert space ${\cal H}$.
However, for describing such states we have to make a change the scale
by letting $\nu$ to infinity. For any $\nu$ the vectors $|\psi_\nu>$
can be represented as vectors of the Hilbert space $L^2(M_\nu)$
of square integrable functions $\psi({\bf q_\nu})$, where $M_\nu$ denotes
the momentum-space corresponding to the scale $\nu$. Obviously, for any
finite $\nu$ the spaces $L^2(M_\nu)$ coincides each with other, since
$\psi({\bf q_\nu})=\psi'({\bf k})=\psi({\bf k}/\nu)$. This, of course, is not 
true for the space $L^2(M_\infty)$ of square integrable functions
$\psi({\bf q}_{in})$, with ${\bf q}_{in}$ being the momentum in the case
when the scale tends to infinity. Since for any $|{\bf k}|<\infty$
$$
\lim\limits_{\nu\tend\infty}<{\bf k}|\psi_{\nu}>=\lim\limits_{\nu\tend\infty}
\int\psi({\bf q}_\nu)<{\bf k}|{\bf q}_\nu>d^3q_\nu=
$$
\begin{equation}
=\lim\limits_{\nu\tend\infty}
\int\psi({\bf q}_\nu)<{\bf q'}_\nu|{\bf q}_\nu>\nu^{-\frac{3}{2}}d^3q_\nu=
\lim\limits_{\nu\tend\infty}
\psi({\bf k}/\nu)\nu^{-\frac{3}{2}}=0,
\end{equation}
each of the vectors $|\psi_{in}>$ of $L^2(M_\infty)$ is orthogonal to 
all vectors of $L^2(M_1)$. 
Thus the Hilbert space describing the states of the system under study is
\begin{equation}
{\cal H}={\cal H}_p\bigoplus{\cal H}_{in},
\end{equation}
where ${\cal H}_p$ is the space that can be realized as the space 
$L^2(M_1)$, and ${\cal H}_{in}$ is the space that can be realized as
the space $L^2(M_\infty)$. The manifold of the physically realizable
states, being the domain of $H_0$, is dense in ${\cal H}_p$. On the other
hand, the vectors of ${\cal H}_{in}$ represent the states with infinite
energy, and are not
physically realizable.

Let us now show that the matrix element $<\psi_{in}^{(2)}|R(t,0)|\psi_{in}
^{(1)}>$, where $R(t,t_0)$ is defined by $U(t,t_0)={\bf 1}+iR(t,t_0)$,
 are not continuous at $t=0$. From (7) and (8) it follows that the operator
$R(t,t_0)$ can be represented in the form
\begin{equation}
R(t,t_0)=-i\int^t_{t_0}dt_2\int_{t_0}^{t_2}dt_1 exp(iH_0t_2)\tilde T(t_2-t_1)
exp(-iH_0t_1).
\end{equation}
By using (53), we can write
$$<\psi_\nu^{(2)}|R(t,0)|\psi_\nu^{(1)}>=-i\int_0^tdt_2\int_0^{t_2}
dt_1\int d^3k'\int d^3k exp(iE_{k'}t_2)\times$$
\begin{equation}
\times exp(-iE_{k}t_1)<\psi_{\nu}^{(2)}|{\bf k}'>
<{\bf k}'|\tilde T(t_2-t_1)|{\bf k}><{\bf k}|\psi_{\nu}^{(1)}>.
\end{equation}
Taking into account (25), (40) and (47), and letting
$\nu$ to infinity, we get
$$
<\psi_{in}^{(2)}|R(t,0)|\psi_{in}^{(1)}>=\lim\limits_{\nu\tend\infty}
<\psi_\nu^{(2)}|R(t,0)|\psi_\nu^{(1)}>=
-ia_1|c_1|^2
\lim\limits_{\nu\tend\infty}\int_0^{t\nu^2} d\theta_2\times
$$
$$
\times\int_0^{\theta_2}
d\theta_1\int d^3q'_\nu\int d^3q_\nu exp(iE_{q'_\nu}\theta_2)
exp(-iE_{q_\nu}\theta_1)\psi_{2}^{*}({\bf q'_\nu})
\psi_{1}({\bf q}_\nu)\times
$$
$$
\times\nu^{-1}|\nu{\bf q}_\nu|^{-\alpha}|{\nu\bf q'_\nu}|^{-\alpha}
\left(\frac{\theta_2-\theta_1}{\nu^2}\right)^{-\frac{1}{2}-\alpha},
$$
where $\theta_i=t_i\nu^2$, $\psi_{i}({\bf q_\nu})=<{\bf q}_\nu|
\psi^{i}_\nu>,$ $i=1,2$.
From this it follows that in the limit $\nu\tend\infty$ the matrix
elements $<\psi_\nu^{(2)}|R(t,0)|\psi_\nu^{(1)}>$ are scale invariant, and
we have
$$
<\psi_{in}^{(2)}|R(t,0)|\psi_{in}^{(1)}>=
-ia_1|c_1|^2
\int_0^\infty d\theta_2\int_0^{\theta_2}
d\theta_1\int d^3q'\int d^3q exp(iE_{q'}\theta_2)\times
$$
\begin{equation}
\times exp(-iE_{q}\theta_1)
\psi_{2}^{*}({\bf q'})
\psi_{1}({\bf q})|{\bf q}|^{-\alpha}|{\bf q'}|^{-\alpha}
(\theta_2-\theta_1)^{-\frac{1}{2}-\alpha}.
\end{equation}
Thus the matrix elements $<\psi_{in}^{(2)}|R(t,0)|\psi_{in}
^{(1)}>$ are independent of $t$, and do not tend to zero as $t\to 0$.
This means that the amplitudes $<\psi_{in}^{(2)}|R(t,0)
|\psi_{in}^{(1)}>$ are not continuous at $t=0$ since, as it follows from 
the definition, $R(0,0)=0$.
From this it follows that the evolution operator (38) is not weakly
continuous. On the other hand one can show that for 
the physically realizable states $|\psi_1>$ and $|\psi_2>$ the matrix
elements $<\psi_2|U(t,0)|\psi_1>$ tend to $<\psi_2|\psi_1>$ as $t\to 0$
and hence the condition (6) is not violated. Nevertheless, the evolution
operator $V(t)=U_s(t,0)$ is not continuous and hence the group of these 
operators has no infinitesimal generator in this case. From this it follows 
that in this case the time evolution of a state vector is not governed by
the Schr{\"o}dinger equation. The cause of this discontinuity is quite
obvious. In fact, from the point of view of the states with infinite
energy any time interval $\delta t$ is infinite and hence the matrix element
$<\psi_{in}^{(2)}|R(\delta t,0)|\psi_{in}^{(1)}>$ must be independent
of $\delta t$, i.e. must be constant. 

The fact that in principle for describing the evolution of a quantum
system during infinitesimal time intervals one has to take into account
the intermediate states with infinite energy is a consequence of the 
principle of uncertainty. In the case of Hamiltonian dynamics, nevertheless,
such states give no contributions because the matrix elements
$<{\bf p}_2|U(t_2,t_1)|{\bf p}_1>$ vanish sufficiently fast when momenta tend
to infinity. In this case the space ${\cal H}_{in}$ can
be ignored, and the dynamics can be restricted to ${\cal H}_p$. 
However, in general case the matrix elements 
$<{\bf p}_2|U(t_2,t_1)|{\bf p}_1>$ 
may have such a large-momentum behavior that one cannot ignore the
space ${\cal H}_{in}$ in describing the dynamics of a quantum system.
To better illustrate this point, let us consider the operator $V_p(t)=
PV(t)P,$ $P$ being the projection operator on the subspace ${\cal H}_p$.
Assume that the group of the operators $V_p(t)$ has a self-adjoint infinitesimal
generator $A$. Then for $|\psi>\in{\cal D}(A)$ we have
\begin{equation}
\frac{V_p(t)|\psi>-|\psi>}{t}\tend\limits_{t\tend 0}-iA|\psi>.
\end{equation}
From this it follows that $A=H_0+A_1$, with
$A_1=-\lim\limits_{t\tend 0}\left (\frac{1}{t} exp(-iH_0t)R(t,0)\right ).$
By using (40) and (53), for $|\psi>\in{\cal D}(H_0)$ and $|{\bf k}|<\infty$,
we get
$$<{\bf k}|A_1|\psi>=i \lim\limits_{t\tend 0}\biggl(\frac{1}{t}
\int_0^tdt_2\int_0^{t_2}
dt_1\int d^3k_1 exp[iE_{k}(t_2-t)]\times
$$
$$
\times
exp(-iE_{k_1}t_1)\varphi^*({\bf k})\varphi({\bf k}_1)
<{\bf k}|\tilde T(t_2-t_1)|{\bf k}_1>\psi({\bf k}_1)\biggr)=$$
\begin{equation}
=ia_1|c_1|^2C\varphi^*({\bf k})\lim\limits_{t\tend 0}\left(\frac{1}{t}
\int_0^tdt_2\int_0^{t_2}dt_1(t_2-t_1)^{-\frac{1}{2}-\alpha}\right)=0,
\end{equation}
where $C=\int d^3k_1\varphi({\bf k}_1)\psi({\bf k}_1)$. Hence $A=H_0$, i.e.
the infinitesimal generator of the group of the operators $V_p(t)$ is 
equal to the free Hamiltonian. This means that the dynamics of the system
cannot be restricted to the space ${\cal H}_p$.

It is important to now whether it is possible to find
the system at time $t$ in a state $|\psi_{in}\in{\cal H}_{in}$,
if initially at time $t_0$ the state of the system was physically realizable.
The amplitude of this possibility is given by the matrix element 
$<\psi_{in}|U(t,0)|\psi_1>$ with $|\psi_1>\in{\cal H}_p$. 
For this matrix element we can write
$$
<\psi_{in}|U(t,0)|\psi_1>=\lim\limits_{\nu\tend\infty}
\int_0^tdt_2\int_0^{t_2}
dt_1\int d^3k'\int d^3k exp(iE_{k}t_2)\times$$
$$
\times exp(-iE_{k'}t_1)\psi^*({\bf k}/\nu)\nu^{-\frac{3}{2}}
\varphi^*({\bf k})<{\bf k}|\tilde T(t_2-t_1)|{\bf k'}>
\varphi({\bf k'})\psi_1({\bf k'})=
$$
$$=a_1c_1^*\lim\limits_{\nu\tend\infty}
\int_0^{t\nu^2} d\theta_2\int_0^{\theta_2}
d\theta_1\int d^3q_\nu\int d^3 k' exp(iE_{q_\nu}\theta_2)\times
$$
\begin{equation}
\times(\theta_2-\theta_1)^{-\frac{1}{2}-\alpha}|{\bf q_\nu}|^{-\alpha}
\psi^{*}({\bf q_\nu})\psi_{1}({\bf k'})\nu^{-\frac{3}{2}}=0.
\end{equation}
Thus, if at time $t_0$ the state of the system was physically 
realizable,
the probability to finding the system at time $t$ in any state $|\psi>\in 
{\cal H}_{in}$  is equal to zero, despite the fact that one cannot ignore the 
subspace ${\cal H}_{in}$. This means that the states $|\psi>\in 
{\cal H}_{in}$ are not observable.
This is not at variance with the fact that one have to take into account
the space ${\cal H}_{in}$ in describing the time evolution of the system.
Indeed, in the limit $t\tend 0$ all matrix elements $<\psi_2|R(t,0)|\psi_1>$ 
tend to zero, provided that $|\psi_1>\in{\cal H}_p$. However, for
$|\psi_2>\in{\cal H}_p$ the matrix elements $<\psi_2|R(t,0)|\psi_1>$ 
vanishes faster than $<\psi_{in}|R(t,0)|\psi_1>$ when $t\tend 0$.
This results in the fact that, as we have seen, the intermediate states
belonging to ${\cal H}_{in}$ are responsible for validity of the composition
law (3) for infinitesimal time intervals. Thus these states play the key
role in the time evolution of the system in the infinitesimal neighborhood
of the point $t=0$.

To clarify this point, note the following. The concept of nonlocal-in-time
potentials were first introduced within the optical-potential model.
The optical potentials are introduced in the case when only state vectors
belonging some subspace of the Hilbert space ${\cal A}$ are included explicitly
in the description of the time evolution of a quantum system. Such
potentials which globally accounts for the coupling between the subspace
${\cal A}$ and its complementary part ${\cal B}$ are nonlocal in time,
and depend on the history of the system. The nonlocal form of the optical
potentials is an expression of the loss of probability from the subspace
${\cal A}$, i.e. of the fact that the evolution operator defined on ${\cal A}$
is not unitary. The nonlocality in time of effective interaction operator
in the case, when
the dynamics of a quantum system is restricted to some subspace of
the Hilbert space, is a consequence of the coupling between this 
subspace and its complementary
part. This clarifies the fact that in our model the interaction operator
becomes nonlocal in time in the case $\alpha<\frac{1}{2}$ when the
ultraviolet behavior of the form factors is "bad", and one cannot ignore
the subspace ${\cal H}_{in}$ of states with infinite energy. In this
case we also have to deal with two subspaces: 
${\cal H}_p$ and its complementary
part ${\cal H}_{in}$. However, in our model the coupling between these
subspaces does not lead to the loss of probability from the subspace
${\cal H}_p:$ The evolution operator $PU(t,t_0)P$, $P$ being the orthogonal
projection on ${\cal H}_p$, is unitary. The remarkable feature of the GQD is
that it allows one to restrict the description of the unitary
evolution of the system to the space ${\cal H}_p$, and, at the same time, 
to take into account the coupling between this space and
${\cal H}_{in}$. In fact, as we have seen, the space ${\cal H}_{in}$ 
manifests itself only
in the behavior of the evolution operator $U(t,t_0)$ in the infinitesimal
neighborhood of the point $t=t_0$. On the other hand, the physical
meaning of the generalized interaction operator $H_{int}(t_2,t_1)$ is
that it determines the evolution
of the system in this neighborhood. In order that $H_{int}(t_2,t_1)$ contain 
all needed dynamical information, in the limit $t_2\tend t_1$
the operator $U_H(t,t_0)=\int_{t_0}^t dt_2
\int_{t_0}^{t_2}dt_1 H_{int}(t_2,t_1)$ must be close enough to the true
evolution operator $U(t,t_0)$.
Thus the solution of the evolution problem is divided in two stages:
The first step is to determine the operator $H_{int}(t_2,t_1)$ describing
the time evolution in the infinitesimal neighborhood of the point
$t=t_0$, where the states belonging to ${\cal H}_{in}$ play an important role,
and the next step is to describe the dynamics restricted to ${\cal H}_p$ starting
with the above generalized interaction operator that takes into account 
the coupling between the spaces ${\cal H}_p$ and ${\cal H}_{in}$.

To illustrate this point, let us come back to our model. From (29) it
follows that the solution of Eq.(27) can be represented in the form
\begin{equation}
t(z)=\lim\limits_{a\tend -\infty}g_a \left(1+g_aJ(z,a)\right)^{-1},
\end{equation}
where
$$
J(z_1,z_2) = (z_1-z_2)  \int d^3k \frac {|\varphi ({\bf k})|^2}
{(z_1-E_k)(z_2-E_k)}.$$
In the case $\alpha <\frac{1}{2}$, $J(z,a)$ tends to infinity like 
$|a|^{\frac{1}{2}-\alpha}$ when $a\tend -\infty$. From this it follows that,
for the solution $t(z)$ to be nonzero, $g_a$
must decrease like $|a|^{\alpha-\frac{1}{2}}$ as $a\tend-\infty$. 
Thus the leader term of
$g_a$ in the limit $a\tend -\infty$ must be of the form $b_1|a|^{-\frac{1}{2}
+\alpha}$, where the parameter $b_1$ is completely determined by the
above requirement of nontriviality of the solution. 
Indeed, by using the identity
$J(z_1,z_2)=J(z_1,z)+J(z,z_2)$, we can rewrite (59) in the form 
\begin{equation}
t(z)=\lim\limits_{a\tend -\infty}g_a \left(1+g_aJ(0,a)+g_aJ(z,0)\right)^{-1}.
\end{equation}
From this formula we see that $g_aJ(0,a)$ must tend to one as $a\tend -\infty$,
since in this limit $g_a$ tends to zero. Hence we have
$$
b_1 =- \lim\limits_{\nu\tend \infty}\left(|a_0|^{\frac{1}{2}+\alpha}
\nu^{2\alpha}\int d^3q_{\nu}
\frac {|\varphi (\nu{\bf q}_{\nu})|^2}
{(E_{q_\nu}-a_0)E_{q_\nu}}\right)^{-1}=
$$
\begin{equation}
=-\left(|a_0|^{\frac{1}{2}+\alpha}|c_1|^{2}\int d^3q_{\nu}
|\psi ({\bf q}_{in})|^2 \right)^{-1},
\end{equation}
where $a=a_0\nu^2$, $a_0\in (-\infty,0)$, $\psi ({\bf q}_{in})=
|{\bf q}_{in}|^{-\alpha}(E_{q_{in}}-a_0)^{-\frac{1}{2}}
E_{q_{in}}^{-\frac{1}{2}}$. Thus the parameter $b_1$ is determined by the
norm of the vector $\psi({\bf q}_{in})$ of the space 
$L^2(M_\infty)$ being the realization of the space ${\cal H}_{in}$. 
From (60) it also follows that the next order term in the asymptotic
behavior of $g_a$ must be of the form $b_2|a|^{-1+2\alpha}$,
where the freedom of choice of the parameter $b_2$ is bounded only by the
requirement of unitarity of the resulting evolution operator. Thus, in the
limit $a \tend -\infty$ the function $t(a)=g_a$ behaves like
$b_1|a|^{-\frac{1}{2}+\alpha}+b_2|a|^{-1+2\alpha},$
and hence the generalized interaction operator must be of the form (36),
and the corresponding solution of Eq.(17) is as follows:
\begin{equation}
<{\bf p}_2|T(z)|{\bf p}_1> =- \varphi^*({\bf p}_2)\varphi({\bf p}_1)
\left(b_2b_1^{-2} + \int d^3k \frac {z|\varphi ({\bf k})|^2}
{(z-E_k)E_k} \right)^{-1}.
\end{equation}
It is easy to verify that (62) is equivalent to the above obtained expression
(32).
Here we are dealing only with the functions belonging to the space
$L^2(M_1)$. Thus after specifying the form of the generalized interaction
operator the description of the dynamics can be restricted to the space
of the physically realizable states. Note that in the case $\alpha>\frac{1}{2}$
such problems do not arise. Indeed, in this case $g_a$ in (29) must tend
to $\lambda$ being an arbitrary real constant. Thus, in this case we directly 
get the expression (32), which is written in terms of functions belonging to
the space $L^2(M_1)$.

By the example of the exactly solvable model we have shown that 
nonlocality in time of the interaction generating the unitary dynamics
of a quantum system should be associated with the existence of the space
${\cal H}_{in}$ of unobservable states that has to be taken into account 
in describing the time evolution of the system. In our model the space
${\cal H}_p$ may be considered as the space of two-nucleon states. The quark 
structure of the nucleons being in such states does not directly manifest
itself since, if the system is in the state $|\psi>\in {\cal H}_p$, the
probability of finding free quarks, when a measurement is performed, is equal
to zero. In this case, the space ${\cal H}_{in}$ may be considered as the
space of states in which the quark structure of nucleons manifests itself.
The spaces ${\cal H}_p$ and ${\cal H}_{in}$ have the same structure.
In the general case the structure of these 
spaces may be different.
For example, in the case of hadron dynamics the space ${\cal H}_{in}$
of states corresponding to infinite momentum scale, i.e. to infinitesimal
length scale,
may describe any states of quarks and gluons, while due to confinement
the space ${\cal H}_{p}$ describes only observable hadron states (color
singlets).
 
Note in this connection that, as we have stated, there are two different
relevant scales in the physics of the strong interaction, and their
separation is so significant that the length scale of processes in which 
quarks and gluons are relevant degrees of freedom can be regarded
as infinitesimal compared to the relevant length scale of the low-energy
hadron physics. Indeed, these degrees of freedom can manifest themselves only
during infinitesimal time intervals. From this in turn it follows that 
the relevant momentum scale of states in which  quarks and gluons
can manifest themselves must be infinitely large. Nevertheless these
degrees of freedom can have significant effects on hadron dynamics, and
for its describing one has to take into account
the coupling between the space of hadron states and the space
 describing states of quark and gluons, and, as usual, this coupling
should result in nonlocality in time of hadron interactions.
However, in contrast with ordinary open quantum systems, due to confinement
this nonlocality must not lead to a loss of probability from the hadron
system: The probability of finding the system in a state containing 
quarks and gluons, when a measurement is performed, is equal to
zero. Nevertheless
one cannot ignore the subspace describing such states. As we have
seen, it is just the case when one has to go beyond Hamiltonian dynamics.
The remarkable feature of the GQD is that it provides the extension of
quantum dynamics to this case. This opens new possibilities
to find the link between the quark-gluon
dynamics and low-energy hadron dynamics. We can try to construct the operator
of hadron interactions by using the quark-gluon dynamics, and then use
it for describing hadron dynamics. As we will show in the next section, for 
constructing this operator one can also use some quark models.

The essential lesson we have learned from the previous analysis is that
the existence of the quark and gluon degrees 
of freedom confined within hadrons gives rise to the peculiar dynamical
situation: In the large-momentum limit the matrix elements of the
evolution operator do not decrease so rapidly as it is required by ordinary
quantum mechanics, and this results in the above-mentioned lack of
continuity of the evolution operator, and correspondingly in the fact
that hadron dynamics is not governed by the Schr{\"o}dinger equation. 
More precisely, the above are retardation effects from quark confinement 
on hadron dynamics, since the existence of some external 
degrees of freedom in itself cannot have such effects on the dynamics of a
quantum system. As we have seen, the retardation effects from quark 
confinement can be very essential. For
example, as we will show in Sec.VI, these effects
give rise to an anomalous off-shell behavior of the two-nucleon amplitudes.

\section*{V. Generalized Interaction Operator and Quark Models}
As we have shown, quark-gluon retardation
must result in nonlocality in time of hadron interactions.
From this it follows that low-energy hadron dynamics should be
governed by the equation of motion (9)
with a nonlocal-in-time generalized interaction operator
describing the history of the hadron system in an infinitesimal
time interval. Here we mean the dynamical system whose states 
correspond to a complete set of hadronic observables.
In the case of low-energy hadron dynamics, 
we can consider as infinitesimal such time intervals, for which 
only the processes being the fundamental at the hadronic level,
for example, the processes being described by the
baryon-baryon-meson and four-baryon vertices, are relevant.
The generalized interaction operator $H_{int}(t_2,t_1)$ represents
the history of the hadron system in infinitesimal time intervals,
during which the quark and gluon degrees of freedom can manifest
themselves, and in principle it should be extracted from QCD. However,
a nonperturbative treatment of QCD is not possible until now, and for
constructing the generalized interaction operator we have to restrict
ourselves to using some quark models.

Let us now show that the dynamics of nucleons with the internal structure
described by a constituent quark model is governed by
the equation of motion (9) with a nonlocal-in-time operator 
$H_{int}(t_2,t_1)$. In constituent quark models nonstrange 
and strange baryons are
considered to be clusters made of three valence quarks. The Hamiltonian 
used in these models has the following form:
\begin{equation}
 H=\sum_i \frac{p_i^2}{2m_q}-T_{c.m.}+V_{con}({\bf {r}}_{ij})+V_{qq}({\bf
{r}}_{ij}), 
\end{equation}
where ${\bf {r}}_{ij}$ represents the interquark distance, $T_{c.m.}$ is
the kinetic energy of the center of mass motion, $V_{qq}({\bf {r}}_{ij})$
is the one-gluon exchange (OGE) potential, and $V_{con}({\bf {r}}_{ij})$ is
the confinement potential
$$
V_{con}({\bf {r}}_{ij})=-a_c{\bf {\Lambda}}_i{\bf {\Lambda}}_j{\bf
{r}}_{ij}^2.  
$$
Here $a_c$ is the confinement strength, and ${\bf {\Lambda}}_i$ are the
SU(3) color matrices.
In the models baryon states are represented by the wave functions
of a three-quark oscillator with the Hamiltonian (63). As is well known,
such models provide a very satisfactory description of the baryon
spectra. In order to take into account the long- and intermediate- range
part of the interaction, in some models in addition to the OGE
interaction, the quarks belonging to different three-quark clusters
interact via scalar and pseudoscalar meson exchange.

For solving the two-nucleon problem within a constituent quark
model, one has to consider the dynamics of the six-quark system.
Due to confinement potential the six-quark system can be only in the 
two-baryon states $|\psi;2B>$ or in the 
baglike states $|\psi;6q>$. However, for large time intervals the 
system can be only in the two-baryon states $|\psi;2B>$. The characteristic
time interval $\tau_c$, during which the system can be in baglike
states, is very small and depends on the confinement length $\lambda_c$.
For the time intervals large compared with $\tau_c$, the matrix elements
$<\psi_2;6q|U(t,t_0)|\psi_1>$ describing the probability to find the system
in the state $|\psi_2;6q>$ at time $t$ are neglectable, and we can write
\begin{equation}
<\psi_2;6q|U(t,t_0)|\psi_1>=0. 
\end{equation}
From this it follows that for large time intervals the operator
$U_{2B}(t,t_0)$ obtained by projecting
$
U_{2B}(t,t_0)=P_{2B}U(t,t_0)P_{2B},$ where $P_{2B}$ is the projection
operator on the subspace ${\cal H}_{2B}$, can be considered to be unitary.

Let us now show that the representation (7) is valid for this operator.
By using the composition law (3) for the evolution operator $U(t,t_0)$,
we can write
\begin{equation}
U(t,t_0)=\prod_{j=1}^{N}U(t_{j},t_{j-1}),
\end{equation}
where $t_j=t_0+j\xi,$ $\xi=(t-t_0)/N,$  $t_N=t.$
Within  Hamiltonian dynamics the evolution operator $U(t_{j+1},t_j)$
has the form
\begin{equation}
U(t_{j},t_{j-1})=exp(iH_0t_{j})exp[-iH(t_{j}-t_{j-1})]exp(-iH_0t_{j-1}).
\end{equation}
In the limit $\xi\tend 0$ we can write
\begin{equation}
U(t_{j},t_{j-1})\approx exp(iH_0t_{j+1})( {\bf 1}-i\xi H)exp(-iH_0t_{j-1}).
\end{equation}
By using (65), the evolution operator can be represented in the form
\begin{equation}
U(t,t_0)=\lim_{\xi\tend 0}\prod_{j=1}^N exp(iH_0t_{j})
( {\bf 1}-i\xi H)exp(-iH_0t_{j-1}).
\end{equation}
It is easy to verify that (68) can be rewritten in the form
\begin{equation}
U(t,t_0)= {\bf 1}-\lim_{\xi\tend 0}\sum_{l=1}^N\left (i\xi
H_I(t_l)+\sum_{k=1}^l\xi^2H_I(t_l)U(t_l,t_k)H_I(t_{k})\right ).
\end{equation}
Taking the limit $\xi\tend 0$, from (69), we get
\begin{equation}
U(t,t_0)= {\bf 1}-\int_{t_0}^tdt_2\int_{t_0}^{t_2}dt_1\left (2i\delta(t_2-t_1)
H_I(t_1)+H_I(t_2)U(t_2,t_1)H_I(t_{1})\right ).
\end{equation}
Thus, within Hamiltonian dynamics, the evolution operator 
can be represented in the form (7) with the following operator 
$\tilde S(t_2,t_1)$:
\begin{equation}
\tilde S(t_2,t_1)=-2i\delta(t_2-t_1)
H_I(t_1)-H_I(t_2)U(t_2,t_1)H_I(t_{1}).
\end{equation}
Correspondingly, for the operator $U_{2B}(t,t_0)$ defined on the two-baryon
space ${\cal H}_{2B}$ we can write
$$<\psi_2,\beta|U_{2B}(t,t_0)|\psi_1,\alpha>=<\psi_2,\beta|\psi_1,\alpha>+$$
\begin{equation}
+\int_{t_0}^tdt_2\int_{t_0}^{t_2}dt_1<\psi_2,\beta|\tilde S_{2B}(t_2,t_1)
|\psi_1,\alpha>,
\end{equation}
where
$$<\psi_2,\beta|\tilde S_{2B}(t_2,t_1)|\psi_1,\alpha>=<\psi_2,\beta|
P_{2B}\tilde S(t_2,t_1)P_{2B}|\psi_1,\alpha>=$$
$$
=-<\psi_2,\beta|P_{2B}exp(iH_0^{(\beta)}t_2)\lbrace
2i\delta(t_2-t_1)H_I^{(\alpha)} +
$$
\begin{equation}
+H_I^{(\beta)}U_{s}(t_2,t_1)H_I^{(\alpha)} \rbrace
exp(-iH_0^{(\alpha)}t_{1})P_{2B}|\psi_1,\alpha>. 
\end{equation}
Here we use the interaction picture in which the interaction in the baryon
clusters is included into the "free" Hamiltonian $H_0^{(\alpha)}$ of the
corresponding two-baryon channel $H=H_0^{(\alpha)}+H_I^{(\alpha)}$, and 
$|\psi,\alpha>$ is the vector belonging to the channel subspaces 
${\cal H}_\alpha$.

We have shown that the operator $U_{2B}(t,t_0)$
can be represented in the form (7), where 
the operator $\tilde S(t_2,t_1)$ is given by (73). 
As has been shown in Ref.[21], for the 
operator $U_{2B}(t,t_0)$ having the form (7) to be unitary
 for any times $t$ and $t_0$, the operator $\tilde S_{2B}(t_2,t_1)$ must
satisfy the equation
\begin{equation}
(t_2-t_1) \tilde S_{2B}(t_2,t_1) = 
\int^{t_2}_{t_1} dt_4 \int^{t_4}_{t_1}dt_3
(t_4-t_3) \tilde S_{2B}(t_2,t_4) \tilde S_{2B}(t_3,t_1).
\end{equation}
However, the operator $U_{2B}(t,t_0)$ can be considered as a unitary
operator only for time intervals $t-t_0$
large compared with $\tau_c$. Correspondingly, the operator $\tilde
S_{2B}(t_2,t_1)$ satisfies the equation (74) only for time intervals
$t_2-t_1>>\tau_c$. On the other hand, due to confinement time
intervals, which are relevant for the description of low-energy nucleon
dynamics, are large compared with $\tau_c$ (here we are dealing with two
separated length scales). From this it follows that
the low-energy dynamics of the two-nucleon system predetermined by a
constituent quark model is governed by Eq.(9). 

Up to now we restricted
ourselves to the consideration of the two-nucleon dynamics. On the other
hand, there are time intervals large, compared with $\tau_c$, that are
much smaller than the time intervals relevant for the many-nucleon
interactions. Hence, for such "infinitesimal" intervals of time the
operator $\tilde S_N(t_2,t_1)$, which is the two-nucleon operator on the
many-nucleon Hilbert space, satisfies Eq.(9). For this reason, the operator
$\tilde S_N(t_2,t_1)$ related to the operator $\tilde S_{2B}(t_2,t_1)$ 
in the ordinary way can be used as the
generalized interaction operator generating the dynamics of the
many-nucleon system
\begin{equation}
H_{int}(t_2,t_1)=\tilde S_{N}(t_2,t_1).
\end{equation}
From (73) it follows that this operator can be represented in the form
(22), where the long- and
intermediate-range parts of $H_I(t)$ present the one-boson-exchange $NN$
potential arising due to the corresponding potential in the six-quark
Hamiltonian, and its short-range part is a potential arising due to the
one-gluon-exchange potential in this Hamiltonian. The nonlocal term
$H_{non}(t_2,t_1)$ describes the contributions from processes in which
during some time intervals retardation from quark confinement can
take place. The contribution from such processes
cannot be neglected, and the effective $NN$ interaction extracted from 
a constituent quark model must contain the nonlocal-in-time term leading to
the above-mentioned dynamical situation. This is an expression of the fact
that there are some effects, related to the short-range $NN$ system, 
that cannot be described in terms of the $NN$ potentials (see, for example, 
[8]).

\section*{VI.Nonlocality in time of the NN interaction and an anomalous 
off-shell behavior of two-nucleon amplitudes.}

The results given in Sec.IV can be used for constructing models which
allows one to take
into account the quark-gluon retardation effects in describing the NN
interaction. As the first step in this direction, we can generalize
the ordinary separable-potential model that is widely used in nuclear
physics.
At the present time, for two-nucleon separable potentials it is
usually used the  Yamaguchi and Tabakin form factors [27,28] which
in spin-triplet and spin-singlet channels are of the form
\begin {equation}
g_Y({\bf p})=\frac{1}{p^2+\beta^2},
\end {equation}
\begin {equation}
g_T({\bf p})=\frac{(p^2+\nu)}{(p^2+\eta)}\frac{(q_c^2-p^2)}{(p^2+\gamma^2)^k},
\end {equation}
where $g_Y(p)$ and $g_T(p)$ are respectively Yamaguchi and Tabakin 
form factors, and $p=|{\bf p}|$.
Such form factors with the parameters given in Ref. [29] have
been recently used by Rupp and Tjon [30] and Adnikari and Tomio [31].
In Refs.[31], for example, the Tabakin form factor has been used 
in one of the nucleon-nucleon spin channels and Yamaguchi in
the other. 

As it follows from the analysis of Sec.IV, the separable potentials cannot
take into account the quark-gluon retardation in describing the NN
interaction, since in this case the NN interaction is instantaneous. 
For the interaction to be nonlocal in time, in the large-momentum
limit the form factor $\varphi({\bf p})$ must behave like (25) with 
$0<\alpha<\frac{1}{2}$, i.e. must have the asymptotic behavior that
in the case of the separable potentials leads to the ultraviolet 
divergences. In this case the generalized interaction operator must be
nonlocal in time, and have the form (36). As we have shown in Sec.IV, the
coupling between the space ${\cal H}_p$ 
of the observable states of the system under
study and the space ${\cal H}_{in}$ 
of unobservable states  manifests itself only in the 
asymptotic behavior of the form factor $\varphi({\bf p})$ in the limit
$|{\bf p}|\tend\infty$, and in the behavior of the $H_{int}^{(s)}(\tau)$
in the infinitesimal neighborhood of the point $\tau=0$. 
Note that in the case of the NN interaction the space ${\cal H}_p$ should be
considered as the space ${\cal H}_N$ of the two-nucleon states, and 
${\cal H}_{in}$ as the space ${\cal H}_q$ of states, in which the quark
structure of nucleons manifests itself.
Since the parameter $a_1$ in (36) is completely determined by the parameters
$\alpha$ and $c_1$ characterizing the asymptotic behavior of the form
factor $\varphi({\bf p})$, we have only three free parameters $\alpha$, $c_1$, 
and $a_2$ responsible for the coupling between the spaces ${\cal H}_N$
and ${\cal H}_q$.
 However, this true only in the case of the rank-one separable
approximation. In this case these three parameters can be determined,
for example, by fitting the NN phase shifts. In a more general case,
when we cannot restrict ourselves to this approximation, we have much
more parameters responsible for the coupling between the spaces
${\cal H}_N$ and ${\cal H}_q$, and these parameters should be extracted
from quark-gluon dynamics.
Let us consider the rank-one separable model of the $NN$ interaction.
Let the form factor $\varphi({\bf {p}})$  in (24) be of the form
\begin{equation}
\varphi({\bf p})=\chi({\bf p})+c_2 g_Y({\bf p}),
\end{equation}
with
$$\chi({\bf p})=\frac{c_1}{(d^2+{\bf p}^2)^{\frac{\alpha}{2}}},
\quad 0<\alpha<\frac{1}{2},$$ where
$d,$ $c_1$ and $c_2$ are some constants.
Since $\alpha <\frac{1}{2}$, the generalized interaction operator must be of
the form (36), and correspondingly the T matrix is given by (37).
To have a bound-state corresponding to the deuteron 
binding energy  $E_d=2.225 MeV$, in the case of the $np$ scattering, 
the following condition must be satisfied:
\begin {equation}
-b_2+b_1E_d^{\frac{1}{2}-\alpha}+b_1^2M(-E_d)=0.
\end {equation}
Taking into account (79) and fitting the $NN$ phase shifts in the 
range 0-350 MeV,
we have obtained the parameters of the interaction operator given by (36).
The results for the S wave phase shifts are shown in
Figs.(1-3). Values for the corresponding constants are given in Table 1.
It can be seen from Figs.(1-3) that
our model yields nucleon-nucleon phase shifts in good 
agreement with experiment despite the fact that the interaction
operator given by (36) is
rank-one separable. The phase shifts in ${}^1S_0$ channels change
sign as well as in the case of Tabakin-type potentials. At the same time,
our model has more freedom in fitting data 
than the separable-potential model. 

In contrast with other phenomenological models, the form of the
interaction operator used in our model takes into account
quark-gluon retardation effects in describing the $NN$ interaction.
The nonlocal interaction operator given by (36) describes the
history of the two-nucleon system in infinitesimal
time intervals during which the quark and gluon degrees of freedom
can manifest themselves. Being exactly solvable, our model can be
used for investigating some effects of quark-gluon retardation
on low-energy nucleon dynamics. As we have shown,
there is the correspondence between the form of the
generalized interaction operator and the large-momentum behavior of the
matrix elements of the evolution operator. In the nonlocal case, these
matrix elements as functions of momenta do not tend to zero at infinity so
rapidly as it is required by ordinary quantum mechanics, and within the
Hamiltonian formalism this leads to the ultraviolet divergences. 
Correspondingly in the nonlocal case the two-nucleon amplitudes
$<{\bf p}_2|T(z)|{\bf p}_1>$ also do not decrease when $
|{\bf p}_i|\tend\infty$ as it is required by Hamiltonian dynamics. Such a
large-momentum behavior of the two-nucleon amplitudes is a consequence of
the fact that in the nonlocal case the operator $H_{non}(t_2,t_1)$ is not
zero and must satisfy Eq.(23). Another
consequence of nonlocality in time of the $NN$ interaction is that for
fixed momenta ${\bf p}_1$ and ${\bf p}_2$ the matrix elements $<{\bf
p}_2|T(z)|{\bf p}_1>$ tend to zero as $z\to -\infty$, while, in the local
case, they tend to $<{\bf p}_2|V|{\bf p}_1>$ in this limit. To illustrate
this, we present in Fig.4 the off-shell behavior $<{\bf p}_2|T(z)|{\bf
p}_1>$ in the limit $z\to -\infty$. Thus, nonlocality in time of the $NN$
interaction caused by quark-gluon retardation effects gives rise to
an anomalous off-shell behavior of the two-nucleon amplitudes. The
off-shell properties of the amplitudes for the ordinary interaction
operator and the operator containing the nonlocal term are qualitatively
different. 
This is true even when the two interaction operators have approximately the
same phase shifts. Such a large variation in the off-shell behavior of the
amplitudes, even when the interaction operators are identical on-shell,
can have significant effects on three- and many-body results [33].
This gives reason to expect that the anomalous off-shell behavior of the
two-nucleon amplitudes can also have significant effects on nucleon matter
properties. From this in turn it follows that the quark and gluon degrees
of freedom can play a significant role in low-energy nucleon dynamics.

\section*{VII. Conclusion}

 As we have shown, the GQD provides the
extension of Hamiltonian dynamics which can describe the evolution
of a quantum system with confined degrees of freedom that cannot be
associated with some observables. 
We have shown that the existence of such degrees of freedom results
in the fact that one has to deal with two infinitely separated
momentum (or length) scales, one of which is relevant for the observable
degrees of freedom, and another is relevant for unobservable ones.
This gives rise to a peculiar dynamical situation: In the large-momentum
limit the matrix elements of the evolution operator do not decrease
so rapidly as it required by ordinary quantum mechanics, and this results
in the lack of continuity of the evolution operator, and correspondingly
in the fact that the evolution of the system is not governed by the
Schr{\"o}dinger equation. The advantage of the GQD consists in the
fact it permits the solution of the evolution problem in this case.
As we have shown, this open new possibilities for describing 
low-energy hadron dynamics. This dynamics can be described in terms of the
hadronic degrees of freedom, while the generalized interaction
operator accounts for the coupling between the space describing hadron
states and the space describing the states, in which the quark 
structure of hadrons manifests itself. This
operator can be extracted from quark-gluon dynamics, or can be constructed
by some quark models. 
This has been illustrated by the example of the dynamics of nucleons 
with the internal structure described by a constituent quark model.
This dynamics has been shown to be governed by the generalized equation of
motion (9) with nonlocal interaction operator $H_{int}(t_2,t_1)$ given by
(75). 

 We have constructed a model which is an extension of the
ordinary separable-potential model to the case where the $NN$ interaction
is nonlocal in time, and the form of the interaction operator takes into
account retardation effects from quark confinement. Despite its simplicity
the model yields the NN phase shifts in good agreement with experiment.
At the same time, the nonlocal interaction operator constructed within
the model can be used for describing the SR part of the NN interaction.

The main conclusion that may be drawn from the present work is that 
quark-gluon retardation can have significant effects on low-energy
hadron dynamics
 We have shown that one of the important consequences of quark-gluon 
retardation effects is the anomalous off-shell behavior of the
two-nucleon amplitudes: The elements of the two-nucleon T matrix as
functions of momenta have the large-momentum behavior which leads to the
ultraviolet divergences in the Hamiltonian formalism. 
Since the off-shell properties of two-particle
amplitudes are crucial for the three- and many-body results, the anomalous
off-shell behavior of the two-nucleon amplitudes can have substantial 
effects on the dynamics of many
nucleon systems, and on the properties of nuclear matter.
Because such an anomalous behavior of the two-nucleon amplitudes takes
place only in the case, when the interaction operator (22) contains the
nonlocal term $H_{non}(t_2,t_1)$ that account for the coupling with the
space describing states, in which the quark structure of hadrons manifests
itself, the above means
that one cannot ignore this short-ranged term in constructing effective
operators of the NN interaction. This is especially important for solving the
many-nucleon problem.

\section*{Acknowledgments}
We would like to thank W. Greiner, I. N. Mishustin and W. Scheid for
helpful discussions and valuable comments. R.Kh.G. would like to
acknowledge the hospitality of Institut f{\"u}r Theoretische Physik der
Justus-Leibig-Universit{\"a}t,Giessen, where part of this work was completed.

\newpage
\section*{References}

\begin{enumerate}

\item[{[1]}]
S. Pal, M. Hanauske, I. Zakout, H. St{\"o}cker, and W. Greiner, Phys. Rev.
C {\bf 60}, 015802 (1999).
\item[{[2]}]
A. Faessler, F. Fernandez, G. L{\"u}beck and K. Shimizu,
Phys. Lett. B {\bf 112}, 201 (1982); Nucl. Phys. {\bf A402}, 555 (1983).
\item[{[3]}]
K. Br{\"a}uer, Amand Faessler, F. Fernandez and K.Shimizu,
Z. Phys. A {\bf 320}, 609 (1985).
\item[{[4]}]
M. Oka and K. Yazaki, Phys. Lett. B 90 (1980) 41; Nucl. Phys. 
{\bf A402}, 477 (1983).
\item[{[5]}]
D.R. Entem, A.I. Machvariani, A. Valcarce, A.J. Buchmann,
Amand Faessler and F. Fernandez, Nucl. Phys. {\bf A602}, 308 (1996).
\item[{[6]}]
V.I. Kukulin, V.N. Pomerantsev, and A. Faessler, Phys. Rev. C {\bf 59}
3021 (1999).
\item[{[7]}]
Yu.A. Simonov, Nucl. Phys. {\bf A463} 231 (1984).
\item[{[8]}]
Fl. Stancu, S. Pepin, and Ya. Glosman, Phys. Rev. C {\bf 56}, 2779 (1997).
\item[{[9]}]
D. Bartz and Fl. Stancu, Phys. Rev. C {\bf 59} 1756 (1999).
\item[{[10]}]
C.-F. Qiao, H.-W. Huang, and K.-T. Chao, Phys. Rev. D {\bf 54}, 2273
(1996); Phys. Rev. D {\bf 60} 094004 (1999).
\item[{[11]}]
R. Machleidt, K. Holinde and Ch. Elster, Phys. Rep. {\bf 149} No.1, 1 (1987).
\item[{[12]}]
A. D. Lahiff and I. R. Afnan, Phys. Rev. C {\bf 56}, 2387 (1997).
\item[{[13]}]
Yu.S. Kalashnikova, I.M.Narodetsky, and V.P.Yurov,  
Yad. Fiz., {\bf 49}, 632 (1989).
\item[{[14]}]
 Yu.A.Simonov,  Phys. Lett. B {\bf 107}, 1 (1981). 
\item[{[15]}]
 A.G. Baryshnikov, L.D.Blokhintsev, I.M.Narodetsky, and D.A.Savin, 
 Yad. Fiz. {\bf 48}, 1273 (1988).
\item[{[16]}]
 A.N. Safronov, Teor. Mat. Fiz. {\bf 89}, 420 (1991); 
 Yad. Fiz., {\bf 57}, 208 (1994).
\item[{[17]}]
 Yu.A. Kuperin, K.A. Makarov, and S.P.Merkuriev, 
Teor.Mat.Fiz., {\bf 75}, 431 (1988); {\bf 76}, 242 (1989).
\item[{[18]}]
 A. Abdurakhmanov and A.L. Zubarev, Z.Phys. {\bf A322}, 523 (1985).
\item[{[19]}]
 M. Orlowski,  Helv.Phys.Acta. {\bf 56}, 1053 (1983).
\item[{[20]}]
B.O. Kerbikov, Yad. Fiz. {\bf 41}, 725 (1985); Teor. Mat. Fiz. {\bf 65},
379 (1985).  
\item[{[21]}]
R.Kh. Gainutdinov, J. Phys. A: Math. Gen. {\bf 32}, 5657 (1999).
\item[{[22]}] 
 R.P. Feynman, Rev. Mod. Phys. {\bf 20}, 367 (1948).
\item[{[23]}]
  R.P. Feynman  and A.R. Hibbs, Quantum Mechanics and Path
  Integrals, (McGraw-Hill, New York, 1965).
\item[{[24]}]
R. Kh. Gainutdinov and  A.A.Mutygullina, 
 Yad. Fiz. {\bf 60}, 938 (1997) [Physics of Atomic Nuclei, 
{\bf 60}, 841 (1997)].        
\item[{[25]}]
R. Kh. Gainutdinov and  A.A.Mutygullina, 
 Yad. Fiz. {\bf 62}, 2061 (1999) [Physics of Atomic Nuclei, 
{\bf 62}, 1905 (1999)].        
\item[{[26]}] 
  R.F.Streater and A.S.Wightman, PCT, 
  Spin and Statistics, And All That,
  (W.A.Benjamin, New York, 1964). 
\item[{[27]}]
Y.Yamaguchi,  Phys. Rev. {\bf 95}, 1635 (1954).  
\item[{[28]}]  
F.Tabakin,  Phys. Rev. {\bf 174}, 1208 (1968).
\item[{[29]}]  
S.K.Adhikari  and L.Tomio, Ann. Phys. (N.Y.) {\bf 235}, 103 (1994).
\item[{[30]}]  
G.Rupp and J.A.Tjon, Phys.Rev. C {\bf 37}, 1729 (1988).
\item[{[31]}]  
S.K.Adhikari  and L.Tomio, Phys.Rev. C {\bf 51}, 70 (1995).
\item[{[32]}]
V. G. J. Stoks, R. A. M. Klomp, M. C. M. Rentmeester, and J. J. de Swart,
Phys. Rev. C {\bf 48}, 792 (1993).
\item[{[33]}] 
R. Machleidt, F. Sammarruca, and Y. Song, Phys. Rev. C {\bf 53}, 1483
(1996). 

\end{enumerate}
\newpage
\bigskip

TABLE 1. The parameters of the interaction operator obtained \\ by fitting 
 the NN date, $\rho=1MeV^{-1}.$
 
\begin{tabular}{|c|c|c|c|c|c|c|}
\hline
partial wave&  $\alpha$ &$\lambda$ & $b\cdot \rho$ & 
$d\cdot \rho$ & $c_1\cdot \rho^{\alpha-1}$ &
$ b_2\cdot \rho^{1-2\alpha}$ \cr
\hline
${}^3S_1(np)$ & 0.499 & $200.2$ &433.8 & 
$766.2$ &0.015 & $7.538\cdot 10^{-4}$  \cr
\hline
${}^1S_0(np)$ & 0.499 & 406.8 & 356.3 & 
$3.651\cdot 10^6$ & 3.086 & $1.779\cdot 10^{-8}$  \cr
\hline
${}^1S_0(pp)$ & 0.499 & 134.7 & 371.7 & 
$6.763\cdot 10^5$ & $0.421$ & $9.564\cdot 10^{-7}$  \cr
\hline
\end{tabular}

\newpage
\begin {center}
\bigskip {{\bf {Figure captions}}}
\end{center}

${\bf {Fig. 1.}}$ Phase shifts (solid line) in the ${}^3S_1$ chanel for
np scattering, compared to the experimental data (points) [32].

${\bf {Fig. 2.}}$ Phase shifts (solid line) in the ${}^1S_0$ chanel for
np scattering, compared to the experimental data (points) [32].

${\bf {Fig. 3.}}$ Phase shifts (solid line) in the ${}^1S_0$ chanel for 
pp scattering, compared to the experimental data (points) [32].

${\bf {Fig. 4.}}$ The off-shell behavior of $<{\bf p}|T(z)|{\bf p}>$ in
the ${}^3S_1$ chanel for np scattering. The solid curves corresponds to
the model with generalized interaction operator (49), compared to the
model with Yamaguchi potential with parameters given in [30] (dashed
line).

\end{document}